\documentclass[english,twocolumn,twocolumn,showpacs,preprintnumbers,amsmath,amssymb,superscriptaddress]{revtex4}
\usepackage[T1]{fontenc}
\usepackage[latin1]{inputenc}
\usepackage{graphicx}

\makeatletter

\makeatletter

\makeatletter

\makeatletter

\makeatletter

\makeatletter

\makeatletter

\makeatletter

\makeatletter

\makeatletter

\makeatletter

\makeatletter

\makeatletter

\makeatletter

\makeatletter

\makeatletter

\makeatletter

\makeatletter

\makeatletter

\makeatletter

\makeatletter

\makeatletter

\@ifundefined{definecolor}
 {\@ifundefined{definecolor}
 {\@ifundefined{definecolor}
 {\usepackage{color}

}{}
}{}
}{}

\usepackage{dcolumn}

\usepackage{bm}

\newcommand{\pt}{ p_{\rm t}}
\newcommand{\ie}{{\it i.e.}}

\def\lsim{\mathrel{\rlap{\lower4pt\hbox{\hskip1pt$\sim$}}
    \raise1pt\hbox{$<$}}}         
\def\gsim{\mathrel{\rlap{\lower4pt\hbox{\hskip1pt$\sim$}}
    \raise1pt\hbox{$>$}}}         

\makeatother

\def\citet{\cite}

\makeatother

\makeatother

\makeatother

\makeatother

\makeatother

\makeatother

\makeatother

\makeatother

\makeatother

\makeatother

\makeatother

\makeatother

\usepackage{babel}
\makeatother
\begin{document}

\title{QGP formation time and direct photons from heavy ion collisions}

\author{Fu-Ming Liu}

\email{liufm@iopp.ccnu.edu.cn}

\author{Sheng-Xu Liu}

\address{Key laboratory of Quark and Lepton Physics (MoE) and Institute of
Particle Physics, Central China Normal University, Wuhan 430079, China}

\date{\today}

\begin{abstract}
We investigated the information carried by the data of direct photons,
$\ie,$ the transverse momentum spectrum and the elliptic flow $v_{2}$
from Pb+Pb collisions at $\sqrt{s_{NN}}$=2.76TeV measured at Large
Hadron Collider (LHC) and from Au+Au collisions at $\sqrt{s_{NN}}$=
200 GeV measured at Relativistic Heavy Ion Collider (RHIC), in the
frame work of (3+1)-dimensional ideal hydrodynamical models constrained
with hadronic data. We found those direct photon data may serve as
a useful clock at the early stage of heavy ion collisions. The time
scales of reaching thermal and chemical equilibrium, extracted from
those data, are about 1/3 and 1.5 fm/c, respectively. Thus the large elliptic flow of direct photons is explainable. High order harmonics,
$\ie$, $v_3$, $v_4$ and $v_5$, of direct photons from Pb+Pb collisions at 2.76TeV
are also predicted, as a further test to compete with those who claim
new sources of photons to account for the large elliptic flow of direct
photons. 
\end{abstract}
\maketitle

\section{Introduction}

Recently, a large elliptic flow of direct photons, as large as that
of hadrons, has been observed in heavy ion collisions, in both PHENIX
experiment at RHIC and ALICE experiment at LHC~\citet{PHENIX2,ALICE2}.
Such a large elliptic flow of direct photons looks puzzling because
the elliptic flow of direct photons was predicted much lower than
that of hadrons\citet{vanHees:2011vb,Chatterjee:2013naa,Chatterjee:2005de,Liu:2009kta}.

New sources of direct photons~\citet{Basar:2012bp, Bzdak:2012fr}
 have been considered
to account for the large elliptic flow. They should, at one hand be
constrained with the observed transverse momentum spectra of direct
photons \citet{PHENIX3,ALICE2} and hadronic data, at the other hand
be tested by higher order harmonics, for example, the triangular flow
$v_{3}$ of direct photons.

In this paper, we will investigate the special information carried
by direct photons, after being constrained by the data of hadrons
in heavy ion experiments. We will try to explained direct photon data
with the delayed formation of the quark gluon plasma(QGP) in the early
stage, following the some early suggestion of two time scales for
thermal and chemical equilibrium \citet{Biro:1993qt,Shuryak92,Shuryak93}.
Thus QGP formation time $\tau_{QGP}$, the moment when the system
reaches both thermal and chemical equilibrium locally, will be extracted.
In order to make the calculation of direct photons constrained with
hadronic data, we will take (3+1)-dimensional ideal hydrodynamical
models \citet{Werner:2012xh,Hirano}, which can give a reasonable
description to hadronic data such as rapidity distribution, transverse
momentum spectra, elliptic flow, etc. More than one hydro models are
employed, in order to make our conclusions more general. During the
whole calculation of direct photons, we will keep the solutions from
those hydro models valid, and keep the equation of states consistent.

The paper is organized as following: after a brief introduction of
calculate approach of direct photons in section 2, the results will
be presented in section 3, then conclusion in section
4.

\section{Calculation approach}

We will consider direct photons from PbPb collisions at 2.76TeV and
AuAu collisions at 200GeV. The sources of direct photons are simplified
as prompt photons and thermal photons, according to ALICE and PHENIX
measurements of direct photons at high transverse momentum \citet{PHENIX3,ALICE2}.
Prompt photons are calculated to the next-to-leading order contribution
in cold nuclear collisions: \begin{eqnarray}
\frac{dN^{{\rm P}}}{dyd^{2}\pt} & = & T_{AB}(b)\sum_{{\displaystyle ab}}\int dx_{a}dx_{b}G_{a}(x_{a},M^{2})G_{b}(x_{b},M^{2})\nonumber \\
 & \times & \frac{\hat{s}}{\pi}\delta(\hat{s}+\hat{t}+\hat{u})[\frac{d\sigma}{d\hat{t}}(ab\rightarrow\gamma+X)\\
 & + & K\sum_{c}\frac{d\sigma}{d\hat{t}}(ab\rightarrow cd)\int dz_{c}\frac{1}{z_{c}^{2}}D_{\gamma/c}(z_{c},Q^{2})],\nonumber \end{eqnarray}
 where thickness function $T_{AB}(b)$, nuclear parton distribution
functions $G(x,M^{2})$ and cross sections, are the same as in previous
work~\citet{Liu:2008eh,Liu:2011dk}. No consideration of energy loss
in fragmentation functions $D_{\gamma/c}(z_{c},Q^{2})$, in order
to compensate the contribution from jet photon conversions. Prompt
photons are supposed to carry vanishing elliptic flow.

The $\pt$ spectrum of thermal photons reads \begin{equation}
\frac{dN^{{\rm T}}}{dyd^{2}\pt}=\int d^{4}x\Gamma(E^{*},T)\label{eq:E*a}\end{equation}
 where $\Gamma(E^{*},T)$ is the photon emission rate at temperature
$T$ and $E^{*}=p^{\mu}u_{\mu}$, $p^{\mu}$ is the four-momentum
of a photon in the lab frame and $u_{\mu}$ is the flow velocity.

The hydrodynamical models \citet{Werner:2012xh,Hirano} provide us
the energy density $\epsilon$ and flow velocity $u_{\mu}$ at each
space-time point of the system for the calculation. We get the temperature
at each space point according to the equation of state $\epsilon=\epsilon(T)$.
We keep those solutions from hydro models valid all the time. What
may be modified is the photon emission rate $\Gamma$, how strong
photons are emitted.

Now we introduce the two time scales in heavy ion collisions, and
discuss how photon emission depends on. The hydrodynamical description
of the heavy ion systems starts from an initial time $\tau_{0}$,
where a local thermal equilibrium has been assumed to solve the hydro
equations. The high energy density at $\tau_{0}$ ensures a partonic
phase. The ratio between quarks and gluons at $\tau_{0}$ can not
be determined by hadronic data. It takes time for the system to get
chemically equilibrium, thus a QGP may form at a later moment $\tau_{QGP}$,
not at $\tau_{0}$.

Quark fugacity $\xi$ is used during ($\tau_{0}$, $\tau_{QGP}$).
A linear increase of quark fugacity $\xi$ from 0 at $\tau_{0}$ to
unity at $\tau_{QGP}$ is assumed in this work. The assumption of
$\xi=0$ until $\tau_{0}$ is at one hand required by the large elliptic
flow of direct photons, at the other hand reasonable from both CGC
and EPOS initial conditions%
\footnote{It is a hot and challenging question in relativistic heavy ion physics
to describe the non-equilibrium system at the early stage. The space-time
evolution of quark fugacity should be determined accordingly. Here
we estimate quark fugacity at $\tau_{0}$ according to initial conditions.
Initial condition can be obtained from Glauber model\citet{Hirano},
according to the distribution of nucleons in nuclei, but difficult
to extract quark fugacity. Color glass condensate (CGC) model provides
initial condition according to parton distribution functions\cite{Gelis:2010nm}
 The small $x$ physics supports a glue-dominant system and $\xi\rightarrow0$.
Event generator EPOS\citet{Werner:2012xh} can provides an initial
condition based on the parallel exchange of Pomerons, a kind of color
tubes with vacuum quantum numbers. The longitudinal excited color
tubes accommodate easily gluons, which implies $\xi\rightarrow0$
at midrapidity.%
}.

Therefore, during ($\tau_{0}$, $\tau_{QGP}$), photon emission rate
is not the full rate in QGP phase $\Gamma_{AMY}$ \citet{AMY2001}.
The Contributions from Compton process and annihilation processes,
$\Gamma_{{\rm Compton}}$ and $\Gamma_{{\rm annihilation}}$ \citet{Kapusta1991},
should be modified with a factor of $\xi$ and $\xi^{2}$, respectively.
Bremsstrahlung process with $n$-quark lines will be modified with
a factor of $\xi^{n}$, where $n\geq2$. It is difficult to disentangle
the contribution of a given $n$-quark Bremsstrahlung. Therefore we
can either ignore Bremsstrahlung contributions to get the lower limit, $\Gamma^{low}$, or overestimate them with $n=2$.
Thus, during ($\tau_{0}$, $\tau_{QGP}$) photon emission rate satisfies
$\Gamma^{low}<\Gamma<\Gamma^{up}$, with the lower limit \begin{equation}
\Gamma^{low}=\xi\cdot\Gamma_{{\rm Compton}}+\xi^{2}\cdot\Gamma_{{\rm annihilation}},\label{eq:rate_s}\end{equation}
 and the upper limit \begin{equation}
\Gamma^{up}=\xi\cdot\Gamma_{{\rm Compton}}+\xi^{2}\cdot(\Gamma_{{\rm AMY}}-\Gamma_{{\rm Compton}}).\label{eq:rate_b}\end{equation}

During this stage, there is also a modification of the relation $\epsilon=\epsilon(T)$
via \begin{equation}
\epsilon=(d_{g}+\xi d_{q})\frac{\pi^{2}}{30}T^{4}\label{eq:et}\end{equation}
 where partonic degrees of freedom $d_{g}=16$ and $d_{q}=31.5$,
a certain increase of the temperature, especially at very small $\xi$.
But the relation between energy density and pressure remains approximately,
so that hydro solutions remain valid.

Now let's summarize photon emission along the whole evolution history:

\begin{itemize}
\item At $\tau=0$, prompt photons are counted according to next to leading
order QCD. 
\item During $0<\tau\leq\tau_{0}$, we have $\xi=0$ and photon emission
rate $\Gamma=0$. 
\item During $\tau_{0}<\tau<\tau_{QGP}$, emission will be estimated with
$\Gamma^{low}<\Gamma<\Gamma^{up}$. 
\item For $\tau\geq\tau_{QGP}$, thermal photon emission rate covers both
contribution from the QGP phase and hadronic phase. In QGP phase,
$\Gamma_{AMY}$ is employed. In hadronic phase, the rate is based
on massive Yang-Mills (MYM) theory\citet{Turbide:2003si},
which takes into account of both nonstrange and strangeness hadronic
interactions such as $\pi+\rho\rightarrow\pi+\gamma$, $\pi+\pi\rightarrow\rho+\gamma$,
$\pi+K^{*}\rightarrow K+\gamma$, etc. 
\end{itemize}
The large elliptic flow of direct photons implies a strong emission
in hadronic phase. Therefore, we don't include the dipole-like form
factor in the emission rate as done in most work\citet{Turbide:2003si,Liu:2008eh}.
This is not only favoured by direct photon data, but also because the form factors
of those hadrons in strong interactions, or PDF of those hadrons,
have not been measured. MYM theory itself remains complete without
form factor.

Now we introduce how to calculate high order harmonics. The $\pt$-spectrum
of thermal photons can be decomposed into harmonics of azimuthal angle
$\phi$ as \begin{equation}
\frac{dN}{d\phi}\sim1+2v_{2}\cos(\phi-\psi_{2})+2v_{3}\cos(\phi-\psi_{3})+...\label{eq:dndphi}\end{equation}
where $v_{2}$ ($v_{n}$) is the elliptic flow (higher order harmonics),
and $\psi_{n}$ is the $n$-th order event plane. Obviously, $v_{n}$
and $\psi_{n}$ depend on photon's transverse momentum $\pt$, and
vary event-by-event. From eq.~(\ref{eq:dndphi}), one can easily
get \begin{eqnarray}
v_{n}\cos n\phi & = & \frac{1}{N}\int_{0}^{2\pi}\cos n\phi\frac{dN}{d\phi}d\phi,\nonumber \\
v_{n}\sin n\phi & = & \frac{1}{N}\int_{0}^{2\pi}\sin n\phi\frac{dN}{d\phi}d\phi.\label{eq:vncs}\end{eqnarray}
Let's note their right sides as $<\cos n\phi>$ and $<\sin n\phi>$,
respectively. Then, in each event one can estimate 

\begin{equation}
v_{n}={\sqrt{<\cos n\phi>^{2}+<\sin n\phi>^{2}}},\label{eq:v2a}\end{equation}
make the event average to get $v_{n}$ of thermal photons and
reduce with the factor $\frac{dN^{{\rm T}}/dp_{t}}{dN^{{\rm T}}/dp_{t}+dN^{{\rm P}}/dp_{t}}$
to get $v_{n}$ of direct photons.

\section{Results}

\subsection{An event-by-event calculation of direct photons from Pb+Pb collisions
at $\sqrt{s_{NN}}=2.76$~TeV}

Let's start with Pb+Pb collisions at $\sqrt{s_{NN}}=2.76$~TeV. Event-by-event
thermal photon emission has been calculated based on EPOS2.17v3\citet{Werner:2012xh}
where the initial time $\tau_{0}=0.35$~fm/c. %
\begin{figure*}
\includegraphics[scale=0.7]{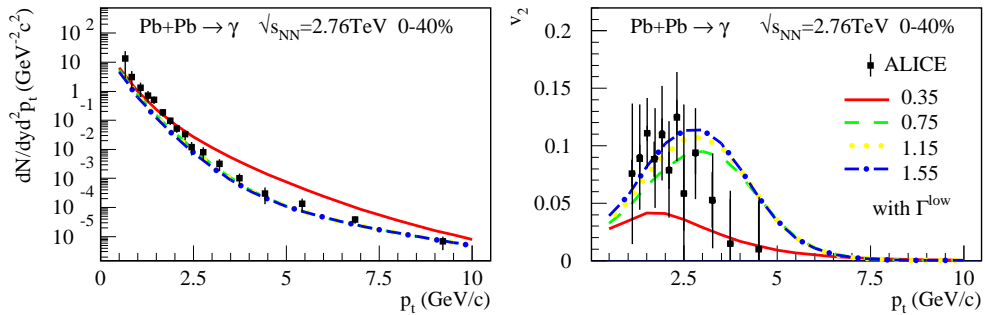}

\caption{\label{fig:Fig1} (Color Online) Transverse momentum spectrum and
elliptic flow $v_{2}$ of direct photons from Pb+Pb collisions at
$\sqrt{s_{NN}}=2.76$~TeV for centrality 0-40\%, calculated with
$\tau_{QGP}=0.35,$ 0.75, 1.15 and 1.55~fm/c. Data points from ALICE
\citet{ALICE2}. }
\end{figure*}

In Fig.\ref{fig:Fig1}, the transverse momentum spectrum and elliptic
flow $v_{2}$ of direct photons from Pb+Pb collisions at centrality
0-40\% calculated with $\tau_{QGP}=0.35,$ 0.75, 1.15 and 1.55~fm/c
are compared with the data from ALICE \citet{ALICE2}. During $\tau_{0}<\tau<\tau_{QGP}$,
the lower emission limit was used in those curves. 

Fig.~\ref{fig:Fig1} tells us, if QGP is formed at the initial time
$\tau_{0}$, then direct photons will be overproduced, and the elliptic
flow underestimated. The underestimation of elliptic flow is consistent
to previous work\citet{vanHees:2011vb,Chatterjee:2013naa,Chatterjee:2005de,Liu:2009kta}. 

A delayed QGP formation will decrease the early photon emission thus
increase the elliptic flow of direct photons, shown in Fig.~1. The
reason is clear. Later emitted photons carry larger elliptic flow,
thanks to a longer expansion of the system. Once we reduce the fraction
of early emission via delayed QGP formation time, the total elliptic
flow will increase. 

The dependence of the transverse momentum spectrum and elliptic flow
$v_{2}$ of direct photons on $\tau_{QGP}$ can be shown more clearly
in Fig.~2. The rate during $\tau_{0}<\tau<\tau_{QGP}$ is not known
explicitly, but the uncertainty is constrained with the upper (dashed
lines) and lower (solid lines) limits. Here $\pt=2.5$GeV/c are chosen
to present the results, because both transverse momentum spectrum
and elliptic flow show sensitivity to $\tau_{QGP}$ here. Besides,
the elliptic flow $v_{2}$ has a peak close to $2.5$GeV/c.

\begin{figure*}
\includegraphics[scale=0.7]{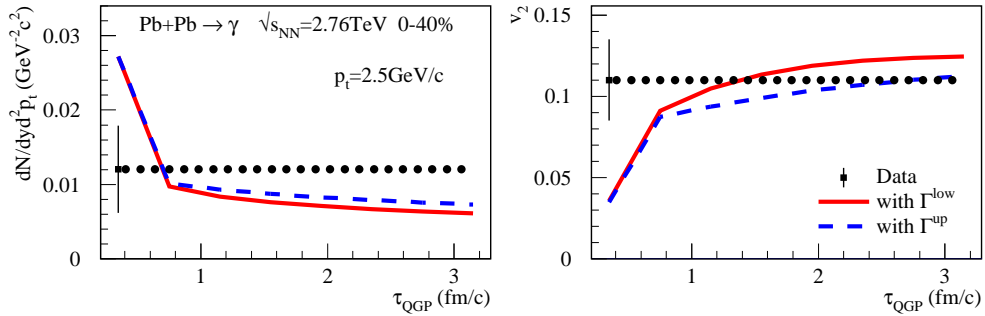}

\caption{\label{fig:Fig2} (Color Online) The dependence of transverse momentum
spectrum and elliptic flow $v_{2}$ on $\tau_{QGP}$ are shown at
$\pt=2.5$GeV, where dashed lines and solid lines represents calculation
with up and low limits respectively between $\tau_{0}$ and $\tau_{QGP}$.
Data points are extracted from ALICE data \citet{ALICE2} and horizontal
lines are used to guide eyes.}
\end{figure*}
We can see a delayed QGP formation can increase the elliptic flow
of direct photons, at the same time, can decrease the transverse momentum
spectrum. Now the direct photon data, extracted from ALICE data \citet{ALICE2}
and shown as full squares, are used to extract the proper QGP formation
time. And a reasonable choice of $\tau_{QGP}$ is about 1.5~fm/c.

The rate during $\tau_{0}<\tau<\tau_{QGP}$ is not known explicitly,
but the uncertainty makes a theoretic error less than 10\% for both
the spectrum and elliptic flow of direct photons. In the following,
we will not mention but use the low limit of emission rate directly.
\begin{figure*}
\includegraphics[scale=0.7]{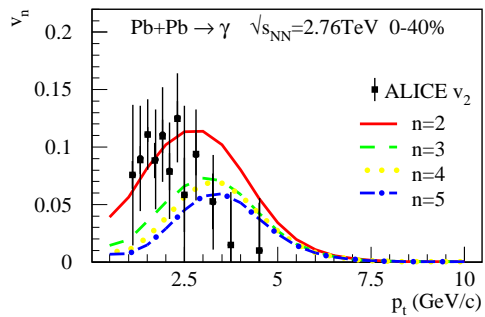}

\caption{\label{fig:Fig3} (Color Online) Predicted harmonics coefficients
$v_{n}$ ($n=2$, 3, 4, 5) of direct photons are various curves. Dots
are measured $v_{2}$\citet{ALICE2}.}
\end{figure*}

In Fig.~3, the harmonics coefficients $v_{n}$ ($n=2$, 3, 4, 5)
of direct photons are predicted with $\tau_{QGP}=1.55$fm/c, accompanied
with the ALICE data points of elliptic flow $v_{2}$. They behave
quite similar to those of charged hadrons measured by ATLAS\citet{ATLAS-vn}.
Thus the two-time-scale picture sounds more reasonable than new sources
of direct photons to account for the large elliptic flow.

\subsection{A connection between event-by-event and event-averaged calculation}

In above case, each system expands hydrodynamically based on an irregular
initial condition and thermal photons are emitted event-by-event.
Here we make a connection to an event-averaged calculation. The latter
has an smoothed initial condition, almond-like in the transverse plane.
This can be obtained from the average of the event-by-event initial
conditions, with $\psi_{2}=0$ in each event, or parameterized with
Glauber model. The regular system expands hydrodynamically and emits
photons, similar to we did previously\citet{Liu:2009kta}.

\begin{figure*}
\includegraphics[scale=0.7]{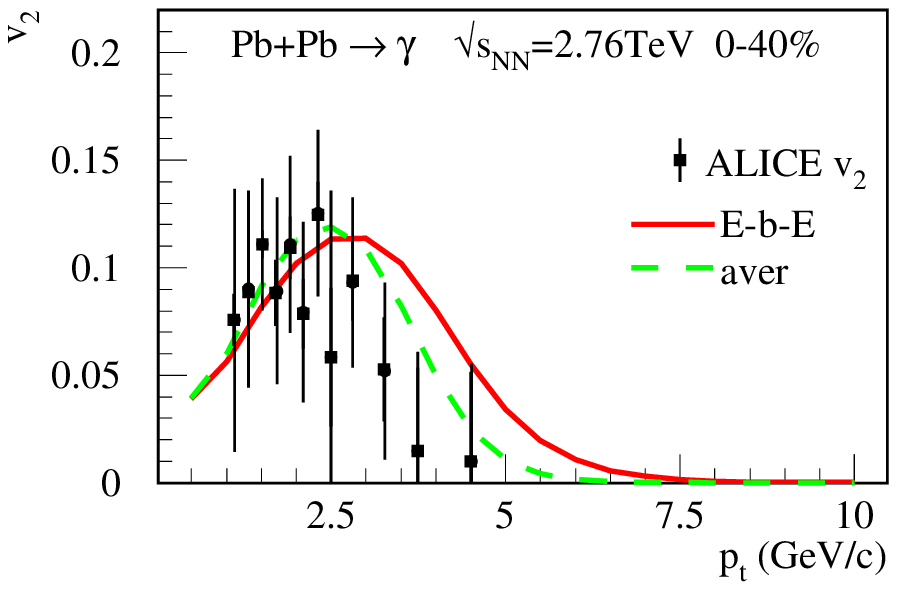}

\caption{\label{fig:Fig4} (Color Online) Elliptic flow of direct photons
based on event-by-event and averaged calculation. Dots are measured
$v_{2}$\citet{ALICE2}.}
\end{figure*}

In Fig.~\ref{fig:Fig4}, the elliptic flow of direct photons from
averaged calculation (dashed line) is compared to event-by-event one
(solid line). The same maximum of elliptic flow are obtained, no matter
event-by-event or averaged calculation. But the high order harmonics
such as $v_{3}$, $v_{4}$and $v_{5}$ vanish in the averaged calculation
because of mixing the irregular events.

The event-by-event curve moves leftward to reach the averaged curve.
Further movement is needed to reach the data shape. As we know viscosity
plays a more important role for irregular systems, we may attribute
the deviation from data shape to the lack of viscosity in the ideal
hydrodynamics.

\subsection{Direct photons from Au+Au collisions at
$\sqrt{s_{NN}}=200$~GeV }

A correct choice of $\tau_{0}$ is also important to get the large
elliptic flow. $\tau_{0}=0.35$~fm/c is provided
directly above. Various $\tau_{0}$ should be checked, but not available
if we requires a good reproduction of hadronic data. One available
case is $\tau_{0}=0.6$~fm/c, as our previous work\citet{Liu:2008eh}
for Au+Au collisions at $\sqrt{s_{NN}}=200$~GeV. This is a (3+1)-dimensional
ideal hydrodynamics\citet{Hirano} with Glauber initial condition.
Averaged EPOS initial condition with $\tau_{0}=0.6$~fm/c provides
the same elliptic flow at midrapidity\citet{Liu:2009kta}, but a different
rapidity dependence. So the following midrapidity discussion is general,
valid for both of the two models. 

\begin{figure*}
\includegraphics[scale=0.7]{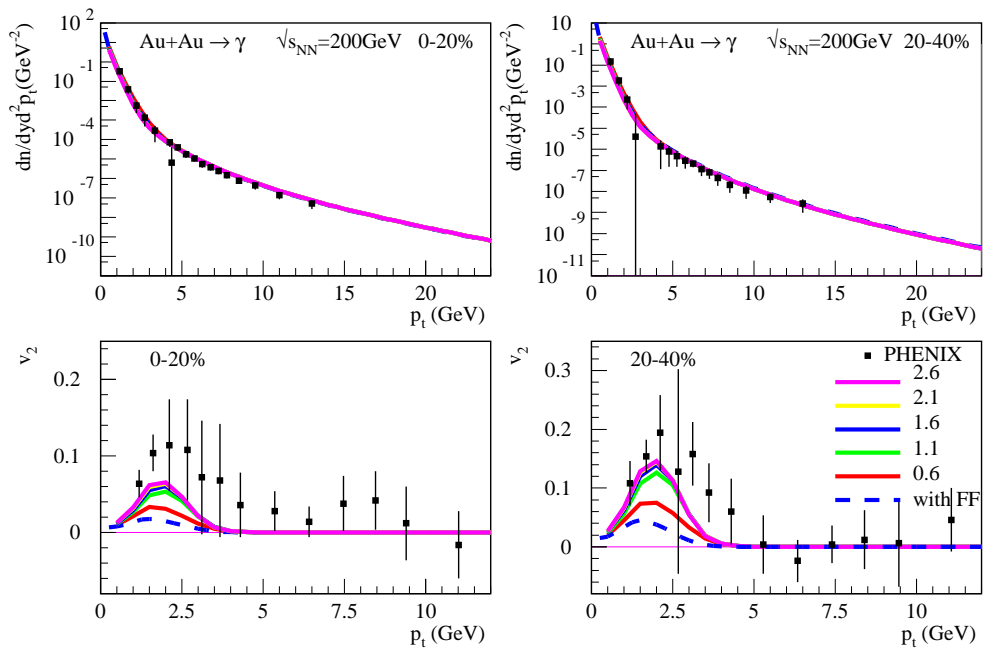}

\caption{\label{fig:Fig5} (Color Online) Transverse momentum spectrum and
elliptic flow $v_{2}$ of direct photons from Au+Au collisions at
$\sqrt{s_{NN}}=200$~GeV for centrality 0-20\% and 20-40\%, calculated
with $\tau_{QGP}=0.6,$ 1.1, 1.6, 2.1 and 2.6 ~fm/c. The spectrum
is not very sensitive to $\tau_{QGP}$. The elliptic flow increases
with $\tau_{QGP}$ then saturate. Data points from PHENIX \citet{PHENIX3}\citet{PHENIX2}. }
\end{figure*}

In Fig.~\ref{fig:Fig5}, the transverse momentum spectrum and elliptic
flow $v_{2}$ of direct photons from Au+Au collisions at $\sqrt{s_{NN}}=200$~GeV
for centrality 0-20\% and 20-40\%, calculated with $\tau_{QGP}=0.6,$
1.1, 1.6, 2.1 and 2.6 ~fm/c are compared with PHENIX data points
from PHENIX \citet{PHENIX3}\citet{PHENIX2}. The previous result
with hadronic form factors\citet{Liu:2008eh} (dashed lines) are very
close to those calculated without form factors (solid lines), because
the spectrum at high $\pt$ is dominant by prompt photons, while at
low $\pt$, form factors are close to unity. 

The solid curves are overlapped
in the upper panels, which shows the insensitivity of spectrum to $\tau_{QGP}$.
In the lower panels, the solid curves from down to up are calculated with
$\tau_{QGP}=0.6,$ 1.1, 1.6, 2.1 and 2.6 ~fm/c, respectively. The
elliptic flow first increases with $\tau_{QGP}$, then saturates.
At $\pt=2$~GeV/c, the maximum of elliptic flow is only 60\% of the
measured value, for both centralities. Thus, $\tau_{0}=0.6$~fm/c
can not work. 

\section{Conclusion and discussion}

The large elliptic flow and the transverse spectrum of direct photons
from Pb+Pb collisions at $\sqrt{s_{NN}}=2.76$~TeV were explained
with $\tau_{0}\sim1/3$~fm/c and $\tau_{QGP}\sim1.5$~fm/c.
High order harmonics coefficient such as $v_{3}$, $v_{4}$ and $v_{5}$
of direct photons were predicted, which behave also quite similar
to those variables of charged hadrons. 

$\tau_{QGP}$ has been studied systematically in this work. The test
of $\tau_{0}$ has been done with two values, 0.35 and 0.6~fm/c in
this work. 

With $\tau_{0}=0.6$~fm/c, the large elliptic flow of direct photons
from Au+Au collisions at $\sqrt{s_{NN}}=200$~GeV can not be fully
reproduced. The delayed QGP formation time can make the elliptic flow
larger, but only up to 60\% of the measured value. 

More work should be done systematically to extract $\tau_{0}$ and
$\tau_{QGP}$ , with systems such as AA, pA and pp. A full explanation
of the data of both charged hadrons and direct photons at both colliders
are expected.

\begin{acknowledgments}
This work is supported by the Natural Science Foundation of China
under the project No.~11275081 and by Program for New Century Excellent
Talents in University (NCET). FML thanks K. Werner and T. Hirano for
providing the hydrodynamical evolution of two collision systems, and
U. Heinz, L. Mclarran, Y. Schutz, E. Shuryak and K.Werner, for very
helpful discussion. 
\end{acknowledgments}


\begin{thebibliography}{10}


\bibitem{PHENIX2} A.~Adare \textit{et al.} {[}PHENIX Collaboration],
 Phys.\ Rev.\ Lett.\ \textbf{109}, 122302 (2012) {[}arXiv:1105.4126
{[}nucl-ex]]. 


\bibitem{ALICE2} D.~Lohner and f.~t.~A.~Collaboration, 
 arXiv:1212.3995 {[}hep-ex]. 




\bibitem{vanHees:2011vb} H.~van Hees, C.~Gale and R.~Rapp, Phys.\ Rev.\ C
\textbf{84}, 054906 (2011) {[}arXiv:1108.2131 {[}hep-ph]].

\bibitem{Chatterjee:2013naa} R.~Chatterjee, H.~Holopainen, I.~Helenius,
T.~Renk and K.~J.~Eskola, 
 arXiv:1305.6443 {[}hep-ph].

\bibitem{Chatterjee:2005de} R.~Chatterjee, E.~S.~Frodermann, U.~W.~Heinz
and D.~K.~Srivastava, 
 Phys.\ Rev.\ Lett.\ \textbf{96}, 202302 (2006) {[}nucl-th/0511079].


\bibitem{Liu:2009kta} F.~-M.~Liu, T.~Hirano, K.~Werner and Y.~Zhu,
 Phys.\ Rev.\ C \textbf{80}, 034905 (2009) {[}arXiv:0902.1303 {[}hep-ph]].

\bibitem{Basar:2012bp} G.~Basar, D.~Kharzeev, D.~Kharzeev and
V.~Skokov, 
 Phys.\ Rev.\ Lett.\ \textbf{109}, 202303 (2012) {[}arXiv:1206.1334
{[}hep-ph]]. 


\bibitem{Bzdak:2012fr} 
  A.~Bzdak and V.~Skokov,
  Phys.\ Rev.\ Lett.\  {\bf 110}, 192301 (2013)
  [arXiv:1208.5502 [hep-ph]].

\bibitem{PHENIX3} S.~Afanasiev \textit{et al.} {[}PHENIX Collaboration],
 Phys.\ Rev.\ Lett.\ \textbf{109}, 152302 (2012) {[}arXiv:1205.5759
{[}nucl-ex]].



\bibitem{Biro:1993qt} T.~S.~Biro, E.~van Doorn, B.~Muller, M.~H.~Thoma
and X.~N.~Wang, 
 Phys.\ Rev.\ C \textbf{48}, 1275 (1993) {[}nucl-th/9303004]. 


\bibitem{Shuryak92}E.~Shuryak, Phys.\ Rev.\ Lett.\ \textbf{68},
3270 (1992) .

\bibitem{Shuryak93}E.~Shuryak and L.~Xiong, Phys.\ Rev.\ Lett.\ \textbf{70},
2241 (1993) .


\bibitem{Werner:2012xh} K.~Werner, I.~.Karpenko, M.~Bleicher,
T.~Pierog and S.~Porteboeuf-Houssais, 
 Phys.\ Rev.\ C \textbf{85}, 064907 (2012) {[}arXiv:1203.5704 {[}nucl-th]].


\bibitem{Hirano} T. Hirano, U. Heinz, D. Kharzeev, R. Lacey, and
Y. Nara, Phys. Lett. B 636, 299 (2006); J. Phys. G 34, S879 (2007);
Phys. Rev. C 77, 044909 (2008).


\bibitem{Liu:2008eh} F.~-M.~Liu, T.~Hirano, K.~Werner and Y.~Zhu,
 Phys.\ Rev.\ C \textbf{79}, 014905 (2009) {[}arXiv:0807.4771 {[}hep-ph]].

\bibitem{Liu:2011dk} F.~-M.~Liu and K.~Werner, 
 Phys.\ Rev.\ Lett.\ \textbf{106}, 242301 (2011) {[}arXiv:1102.1052
{[}hep-ph]]. 


\bibitem{AMY2001}P.~Arnold, G.~D.~Moore, and L.~G.~Yaffe, J.~High
Energy Phys. \textbf{0111}, 057 (2001); J. High Energy Phys. \textbf{0112},
9 (2001).


\bibitem{Kapusta1991}J.~Kapusta, P.~Lichard and D.~Seibert, Phys.\ Rev.
\ D \textbf{44}:2774,(1991);\textbf{47}:4171(E),(1991).

 

\bibitem{Turbide:2003si} S.~Turbide, R.~Rapp and C.~Gale, 
 Phys.\ Rev.\ C \textbf{69}, 014903 (2004) {[}hep-ph/0308085]. 


\bibitem{ATLAS-vn}ATLAS collaboration, Phys. Rev.C86,014907, (2012). 

\bibitem{Gelis:2010nm} 
  F.~Gelis, E.~Iancu, J.~Jalilian-Marian and R.~Venugopalan,
  Ann.\ Rev.\ Nucl.\ Part.\ Sci.\  {\bf 60}, 463 (2010)
  [arXiv:1002.0333 [hep-ph]].

\end{thebibliography}
\end{document}